\documentclass[12pt]{article}
\usepackage{amsmath}
\newcommand{\bt}{\begin{theorem}}
\newcommand{\ds}{\displaystyle}

\newcommand{\et}{\end{theorem}}
\newcommand{\bea}{\begin{eqnarray}}
\newcommand{\eea}{\end{eqnarray}}

\def \spec#1 {\mathop{#1}}

\def \b #1 {\bf #1}
\newcommand {\be}{\begin{equation}}
\newcommand {\ee}{\end{equation}}

\newcommand{\vsp}{\vskip 1em}

\newcommand{\hsp}{\hskip 2em}

\newcommand{\ben}{\begin{eqnarray*}}
\newcommand{\een}{\end{eqnarray*}}

\def \qed {\hfill \vrule height6pt width6pt depth0pt}
\newtheorem{lemma}{Lemma}
\newtheorem{theorem}[lemma]{Theorem}

\def \noi{\noindent}

\def \geq{\ge}
\def \leq{\le}
\newcommand{\bdsc}{\begin{description}}
\newcommand{\edsc}{\end{description}}

\def\ba{\begin{array}}
\def\ea{\end{array}}

\title{Strategic delegation in a sequential model with multiple stages}
\begin{document}
\date{}

\author{Paraskevas V. Lekeas{\footnote{Department of Applied Mathematics, University of
Crete, Heraklion, Crete, Greece; email:
plekeas@tem.uoc.gr\vspace{0.2cm}}}\hsp Giorgos
Stamatopoulos{\footnote{Department of Economics, University of
Crete, 74100 Rethymno, Crete, Greece; email:
gstamato@econ.soc.uoc.gr\vspace{0.2cm}}}}

\maketitle
\begin{abstract}\noi We analyze strategic delegation in a
Stackelberg model with an arbitrary number, $n,$ of firms. We show
that $n-1$ firms delegate their production decisions and only one firm (the one whose manager is the first mover) does not. The later a manager commits to a quantity, the higher his incentive rate. Letting $u_i^*$
denote the equilibrium payoff of the firm whose manager commits in the $i$-th
stage, we show that $u_n^*>u_{n-1}^*>\cdots>u_2^*>u_1^*.$ We
also compare the delegation outcome of our game with that of a corresponding
Cournot oligopoly and show that managers who commit late (early) are given higher (lower) incentive rates than managers in the Cournot market.

\end{abstract}

\vspace{0.2cm} \noi\hspace{0.83cm} {\small{\emph{Keywords}:
Sequential competition; late-movers' advantage; delegation

%\vspace{0.2cm} \hspace{0.33cm}\noi \emph{JEL Classification}: D43,
%L13, L21.}}

\section{Introduction}
The Stackelberg model of market competition is a benchmark model of
industrial economics. In this model, firms select their market
strategies (quantities or prices) sequentially.  One of the most important
issues in this framework focuses on the relation between timing of commitment and relative
profitability of firms. For the case of two players, Gal-Or (1985)
showed that if reaction functions are downwards-sloping then the first-mover earns a
higher payoff than his opponent. On the other hand, in the case of
upwards-sloping reaction functions the advantage is with the
second-mover.{\footnote{The result of Gal-Or (1985) is obtained in a
set-up which includes the Stackelberg duopoly as a special case.}} Further studies
showed that this result is not robust to variations of the model.
Gal-Or (1987) studied a Stackelberg duopoly where firms compete under
private information about market demand. In this model the
first-mover might earn a lower profit than his opponent, as he
produces a relatively low quantity in order to send a signal for low
demand. Liu (2005) analyzed a model where only the first-mover has
incomplete information about the demand and showed that in some
cases the first-mover loses the advantage.

For the case of $n\geq 2$ symmetric  firms, Boyer and Moreaux (1986) and
Anderson and Engers (1992) showed that the $i$-th mover  obtains a
higher profit than the $i+1$-th mover, for $i=1,2,\cdots,n-1 $. Pal and
Sarkar (2001) analyzed a model with $n\geq 2$ cost-asymmetric firms
under the assumption that the later a firm commits to a quantity, the
lower its marginal cost. They showed that if cost differentials are
sufficiently low, the firm that moves in stage $i$ obtains a higher
payoff than its successor $i+1$; otherwise, the ranking of profits
is reversed.

Recently, an integration of the Stackelberg model with the theory of
endogenous objectives of oligopolistic firms has taken place. The
latter theory was launched with the works of Fershtman and Judd
(1985), Vickers (1985) and Sklivas (1987). These works endogenized
the objective functions of firms in a context of
management/ownership separation by postulating that firms maximize a
combination of revenue and profit or quantity and profit. This
framework was applied by Kopel and Loffler (2008) to a Stackelberg
duopoly with homogeneous commodities (which give rise to downwards
sloping reaction functions). Their paper analyzed the impact of
delegation on the structure of first versus
second mover advantage. The authors showed that only the second mover
delegates the production decision to a manager. As a
result, the second mover produces a higher quantity than the first mover and earns higher profit.

The current paper analyzes strategic delegation in a Stackelberg model with
an arbitrary number of firms. It assumes a fixed order of play and perfect observability of choices at each stage.{\footnote{The dropping of the last assumption is not without consequences. We disuss this at a later point.}} Our work is an extension of the strategic delegation setup presented in Kopel and Loffler (2008).{\footnote{Kopel and Loffler also considered investment in R$\&$D, but the present paper focuses on their
delegation setup in an $n$-firm oligopoly.}} Our aim is to determine the relations
among: (i) the timing of commitment to quantities; (ii) the equilibrium
delegation decisions and (iii) the relative performance of firms.
Moreover, we are interested in comparing the equilibrium of the sequential market with that of a
corresponding Cournot market.

The main results of the paper are as follows: First, we
show that all firms delegate their production decision to managers except for the
firm whose manager is the first to commit to a quantity. Moreover, the equilibrium incentive rate is an increasing function of the order of commitment. Namely, the later a manager selects a quantity,
the higher his incentive rate he is given. More importantly, letting
$u_i^*$ denote the equilibrium payoff of the firm whose manager commits in stage $i$, we show that $u_n^*>u_{n-1}^*>\cdots>u_2^*>u_1^*.$ This ordering of profits is
due to the result that the managers who commit at late stages choose relatively high quantities (as they are given relatively high incentive rates).

Delegation in a Cournot model leads to an equilibrium where all
firms end-up with a lower payoff compared to the case of
non-delegation. This is not true though for the Stackelberg model:
Firms whose managers decide on quantities after a threshold stage prefer the delegation regime over nondelegation. Nonetheless, we show that if the number of firms is $n\geq 3$, each firm
in the Stackelberg market earns a lower payoff than a Cournot firm.

The rest of the paper is organized as follows. Section 2 describes the model and section 3 presents the results. Section 4 concludes.

\section{The framework}
Consider an $n$-firm sequential oligopoly. Firms face the inverse
demand function $P=max\{a-Q,0\}$, where $P$ is the market
price and $Q$ is the total market quantity given by
$Q=q_1+q_2+...+q_n$, where $q_i$ is the quantity of firm $i=1,2,\cdots,n$. The
production technology of firm $i$ is represented by the cost
function $C(q_i)=cq_i$, $i=1,2,\cdots,n.$ Firms are characterized by ownership-management separation. The task of firm $i$'s manager is to select a
quantity by maximizing an objective function delegated to
him by the owners of the firm. We assume that this objective
function is a combination of profit and quantity (Vickers
1985),{\footnote{The results of this paper do not change if we
assume that the objective function of each firm is a convex
combination of profit and revenue (Fershtman and Judd 1985, Sklivas
1987).}}

$$T_i=(P-c)q_i+a_iq_i, \hspace{0.2cm} a_i\geq 0, \hspace{0.2cm} i=1,2,\cdots,n$$

\noindent where $a_i$ is manager $i$'s incentive rate. The time structure of the interaction among firms and managers is as follows. In stage 0, the firms' owners decide simultaneously
on the incentive rates of their managers. In particular, firm $i$'s owners choose $a_i$ so as to maximize the profit function
$$u_i=(a-Q-c)q_i, \hspace{0.2cm} i=1,2,\cdots,n$$ These choices are made publicly known.
Then play becomes sequential. In stage 1 the manager of firm 1 selects (and commits to) a quantity for his firm. His choice is observed by all other players. In stage 2, firm 2's manager selects a quantity, which is observed by all other players. The process continues this way in stages $3,4,\cdots,n-1,n$.

We denote the above interaction by $G_S$. In the next section we identify the sub-game perfect Nash equilibrium (SPNE) outcome of this game.

\section{Results}
\subsection{Quantity stages} Working backwards, we first analyze the quantity competition stages of $G_s$. We first note that, in essence, managers choose quantities as if their firms face assymetric marginal costs given by $(c_1,c_2,\cdots,c_n) = (c-a_1,c-a_2,\cdots,c-a_n)$. Thus depending on the 0-stage choices of $(a_1,a_2,\cdots,a_n)$ and the resulting asymmetries we can have, {\em a priori}, some managers selecting zero quantities.
We will show though that any configuration with one ore more managers selecting zero quantities cannot be part of a \emph{SPNE} outcome of $G_S$.

We begin by describing the managers' reaction functions. It will be useful for our analysis to define not only the standard reaction function but also the auxiliary concept of {\em step k reaction function} where $k$ is a positive integer) on which we elaborate below.

Let $Q^i= q_1+q_2+\cdots+q_{i-2}+q_{i-1}$. Consider first stage $n$. We will denote by $f_n^1(q_1,...,q_{n-1})$ the {\em step 1 reaction function} or simply the {\em reaction function} of manager $n$, defined by{\footnote{Whenever there is no confusion, we will drop the variables $q_1$, $q_2$, etc., from the definitions of the various reaction functions. For notational simplicity, the definitions do not include the incentive rates.}}
 
\[f_n^1(q_1,\cdots,q_{n-1})= \operatorname*{arg\,max}_{q_n\geq 0}T_n(q_1,\cdots,q_n)\]

\noindent where $T_n(q_1,\cdots,q_n)=(a-Q-c+a_n)q_n$. For the moment we do not discuss the positiveness or not of the reaction functions; we will turn to this (critical) issue later on in the analysis. Moving to stage $n-1$, the (step 1) 
reaction function of manager $n-1$ is $$f_{n-1}^1(q_1,\cdots,q_{n-2})=\operatorname*{arg\,max}_{q_{n-1}\geq 0}T_{n-1}(q_1,\cdots,q_{n-1})$$

where $$T_{n-1}(q_1,\cdots,q_{n-1})=(a-Q^{n-1}-q_{n-1}-f^{1}_n-c+a_{n-1})q_{n-1}$$

\vspace{0.3cm}

\noi Then the {\em step 2 reaction function} of manager $n$ is derived by $f_n^1$ when $q_{n-1}$ is replaced by $f_{n-1}^1$, i.e., 
$$f_{n}^2(q_1,\cdots,q_{n-2})=f_n^1|_{\tiny q_{n-1}=f_{n-1}^1}$$

\vspace{0.3cm}\noi Moving on to stage $n-2$, the step 1 reaction function of manager $n-2$ is defined by
$$f_{n-2}^1(q_1,\cdots,q_{n-3})=\operatorname*{arg\,max}_{q_{n-2}\geq 0}T_{n-2}(q_1,\cdots,q_{n-2})$$

where $$T_{n-2}(q_1,\cdots,q_{n-2})=(a-Q^{n-2}-q_{n-2}-f_{n-1}^1-f_n^2-c+a_{n-2})q_{n-2}$$

\vspace{0.3cm}

\noi Plugging $f_{n-2}^1$ into $f_{n-1}^1$ and $f^2_n$ will give us the step 2 reaction function of manager $n-1$ and the {\em step 3 reaction function} of manager $n$. These are respectively,
 $$f_{n-1}^2(q_1,...,q_{n-3})=f_{n-1}^1|_{\tiny q_{n-2}=f_{n-2}^1}, \; f_n^3(q_1,\cdots,q_{n-3})=f_n^2|_{\tiny q_{n-2}=f_{n-2}^1}$$

\vspace{0.3cm}\noi We can iteratively continue this way and define the reaction functions up to stage 2. Then, in stage 1, manager 1 solves $max_{q_1\geq 0}T_1(q_1)$ where

$$T_1(q_1)=(a-q_1-\sum\limits_{k=2}^{n}f_k^{k-1}-c+a_1)q_1$$

\vspace{0.3cm}\noi Let $q_1^*$ denote manager 1's choice. Then  the choice of manager 2 is $q_2^*=f_2^1(q_1^*)$; of manager 3 is $q_3^*=f_3^2(q_1^*)=f_3^1(q_1^*,q_2^*)$ up to manager $n$'s choice which is $q_n^*=f_n^{n-1}(q_1^*)=\cdots=f_n^1(q_1^*,\cdots,q_{n-1}^*)$.

\noindent The above description will be useful in order to examine what type of quantity configurations can support an SPNE outcome of $G_S$. To this end, consider the generic stage $i$. Using our description, manager $i$ selects $q_i$ in order to maximize the function{\footnote{To be consistent, when dealing with $T_1$ we need to set $Q^1=0$.}} 

$$T_i(q_1,\cdots,q_1)=(a-Q^i-q_i-\sum\limits_{k=i+1}^{n}f_{k}^{k-i}-c+a_i)q_i$$

Notice that

\vspace{0.3cm}$$\frac{\partial{T_i(q_1,\cdots,q_i)}}{\partial{q}_i}=
a-Q^i-2q_i-\sum\limits_{k=i+1}^{n}(f_{k}^{k-i}-\frac{\partial{f_k^{k-i}}}{\partial{q}_i}q_i)-c+a_i$$.

\vspace{0.3cm}\noi Recall that $f^1_i(q_1,\cdots,q_{i-1})$ denotes manager $i$'s (step 1) reaction function. Then the concavity of $T_i$ in $q_i$ (which can be easily established) implies that if ${\ds{\frac{\partial{T_i(q_1,\cdots,0)}}{\partial{q}_i}\leq 0}}$ then $f_i^1(q_1,\cdots,q_{i-1})=0$ whereas if
${\ds{\frac{\partial{T_i(q_1,\cdots,0)}}{\partial{q}_i}>0}}$ then $f_i^1(q_1,\cdots,q_{i-1})>0.$
These conditions in turn imply that: 

\emph{(i)} if $Q^i+\sum_{k=i+1}^n f^{k-i}_{k}(q_1,\cdots,0)\geq a-c+a_i$ then $f_i^1(q_1,\cdots,q_{i-1})=0$ and

\emph{(ii)} if $Q^i+\sum_{k=i+1}^n f^{k-i}_{k}(q_1,\cdots,0)<a-c+a_i$ then $f_i^1(q_1,\cdots,q_{i-1})>0.$

\vsp\noi We argue that case (i) cannot be part of any SPNE outcome. To this end,
consider a vector $(\tilde{a}_1,\tilde{a}_2,\cdots,\tilde{a}_n)$ of 0-stage choices. Assume that these choices are such that all managers select positive quantities except for one, say manager $i$. Let $(\tilde{q}_1,\cdots,\tilde{q}_{i-1},0,\tilde{q}_{i+1},\cdots,\tilde{q}_n)$
denote this market outcome. Given that $\tilde{q}_k=f^{k-i}_k(\tilde{q}_1,\cdots,\tilde{q}_{i-1},0)$, $k=i+1,i+2,\cdots,n$, and since we are in case (i), we have $\sum_{j \neq i} \tilde{q}_j \geq a-c+\tilde{a}_i$. But then the profit of any firm $j, j\neq i,$ in stage 0 is

$$u_j=(a-\tilde{q}_1-\cdots-\tilde{q}_n-c)\tilde{q}_j\leq (a-(a-c+\tilde{a}_i)-c)\tilde{q}_j=-\tilde{a}_i\tilde{q}_j\leq 0$$

\vspace{0.2cm}\noi To put it differently, a configuration of the form $(\tilde{q}_1,\cdots,\tilde{q}_{i-1},0,\tilde{q}_{i+1},\cdots,\tilde{q}_n)$ cannot support an SPNE outcome as such a configuration would make the market price fall bellow the marginal cost. To see this, notice that the conditions $\sum_{j \neq i}\tilde{q}_j \geq a-c+\tilde{a}_i$ and $\tilde{a}_i \geq 0$ imply that $a-\sum_{j \neq i}\tilde{q}_j \leq c$. A similar argument holds for outcomes under which more than one managers select zero quantities. Hence in what follows we can focus our attention on the case where all firms produce positive quantities.

Since in any SPNE outcome all managers produce positive quantities,
we can use the results of Pal and Sarkar (2001) who computed the equilibrium
quantities in an $n$-stage Stackelberg with cost-asymmetric firms
(but without delegation) under the assumption that all firms are active.
By adjusting their analysis to ours, the manager of firm $i$ chooses the
quantity 

\be
q_i^*=(P^*-c+a_i)2^{n-i}, \hsp i=1,2,\cdots,n
\ee 

where

$${\ds{P^*=\frac{a}{2^n}+\sum_{i=1}^n\frac{c-a_i}{2^i}}}$$ 

is the market price.

\subsection{Delegation stage}
Armed with the above, we can move to stage 0 (the delegation stage). Let $(a_1,a_2,\cdots,a_n)=(a_i,a_{-i}).$ Using (1), the payoff of firm $i$ in stage 0 is
$$u_i(a_i,a_{-i})=2^{n-i}(\frac{a}{2^n}+\sum_{j=1}^n\frac{c-a_j}{2^j}-c)
(\frac{a}{2^n}+\sum_{j=1}^n\frac{c-a_j}{2^j}-c+a_i)$$ The
maximization problem facing firm $i$'s owners is $max_{a_i}u_i(a_i,a_{-i})$,
$i=1,2,\cdots,n$. Define $D_i=2^{i+1}/(\sigma(i)-1)$,
$\sigma(i)=(2^{i+1}-2)/(2^i-2)$ and $h(n)=-2+2n+2^{2-n}.$

\vsp\noi \textbf{Lemma 1.} \emph{Consider the delegation stage of $G_S$.}

\vspace{0.2cm}\noi\emph{(i) The equilibrium incentive rates are
$$a_1^*=0, \hspace{0.3cm} a_i^*=D_i\frac{a-c}{2^nh(n)}>0,\hspace{0.2cm}i=2,3,\cdots,n$$}
\noi\emph{(ii) \emph{The inequalities} $a_n^*>a_{n-1}^*>\cdots>a_2^*>a_1^*$ \emph{hold}.}

\vspace{0.3cm}\noi \textbf{Proof.} Appears in the Appendix.

\vsp\noi By Lemma 1 all firms except for 1, delegate in equilibrium. Moreover, the later a manager commits to a quantity, the higher his incentive rate. To give some intuition behind this result,
we first note that there is a negative relation between any $a_i$ and the market price. For example, consider firms $i$ and $i+j$ with $j>0$. The corresponding effects of $a_i$ and $a_{i+j}$ on the market prices are

$${\ds{\frac{\partial{P^*}}{\partial{a_i}}=-\frac{1}{2^i}<\frac{\partial{P^*}}{\partial{a_{i+j}}}}}=-\frac{1}{2^{i+j}}.$$

\noindent To comprehend why the above inequality holds, let us go back to the managers' (step 1) reaction functions: the rate $a_i$ appears in the step 1 reaction functions\footnote{Recall by the previous footnote that for reasons of notational simplicity we have not included the incentive rates within the definitions of the reaction functions.} of $q_1,\cdots,q_{i-1},q_i$ whereas the rate $a_{i+j}$ appears in the step 1 reaction of more terms, i.e. $q_1,\cdots,q_i,\cdots,q_{i+j-1},q_{i+j}$. Furthermore: (i) the relation between $a_i$ and any of $q_1,\cdots,q_{i-1}$ is negative and so is the relation between $a_{i+j}$ and any of $q_1,\cdots,q_i,\cdots,q_{i+j-1}$; (ii) the market price depends negatively on quantities. Points (i) and (ii) explain why $a_{i+j}$ has a smaller negative impact on price than $a_i$ has. As a result, the owners of firm $i+j$ have incentive to make their manager more aggressive than firm $i$'s owners.

\noindent Using Lemma 1, market price, individual and total market quantities are
given respectively by 

\be P_S^*=\frac{a}{2^{n-1}h(n)}+\frac{c(1+n2^n-2^{-n})}{2^{n-1}h(n)}\ee

\be q^*_{iS}=(2-2^{1-i})\frac{a-c}{h(n)}, \hspace{0.3cm}\noi
i=1,2,\cdots,n\ee

\be{\ds{Q_{S}^*=(a-c)[1-2^{-n}+\frac{2n-4+2^{2-n}}{2^nh(n)}]}}\ee

\vsp\noi Let $u_i^*$ denote the equilibrium profit of firm $i,
i=1,2,\cdots,n$, in $G_S$. Our next result ranks these profits.

\vsp\noi \textbf{Proposition 1.} \emph{The inequalities
$u_n^*>u_{n-1}^*>\cdots>u_2^*>u_1^*$ hold in $G_S$.}

\vspace{0.2cm}\noi \textbf{Proof.} Since firms face the same price
and they are cost-symmetric, $u_{i+1}^*>u_i^*$ if and only if
$q_{i+1}^*>q_i^*$, $i=1,2,\cdots,n-1,$ which holds by inspection of (3). \qed

\vsp\noi One question raised at this point is how does the
performance of firms in $G_S$ compare with their performance
in a sequential market without any delegation activities.
Let $\bar{u}_i$ denote the equilibrium profit of the $i$-th
firm in the latter market. We have the following.

\vsp\noi \textbf{Corollary 1.} \emph{There exists a stage $i'=i'(n)$
such that $u_i^*>\bar{u}_i$ if and only if
 $i>i'(n)$.}

\vspace{0.2cm}\noi \textbf{Proof}. Appears in the Appendix

\vspace{0.2cm}

\noindent Therefore, firms whose managers select quantities after the i'th stage prefer the delegation regime over nondelegation; the opposite holds for the remaining firms. This result is explained by our previous finding that the late-moving managers are relatively aggressive at the expense of the early movers.

\subsection{Comparison with Cournot competition}
In this section we compare the equilibrium outcome of $G_S$ with the
outcome of the corresponding Cournot market. In the latter
framework, we have a two stage interaction which evolves as follows: in stage 0, firms' owners choose
the incentive rates of their managers. These choices are made publicly known. Then in stage 1,
the managers of the $n$ firms select simultaneously quantities for
their firms, using the incentive schemes decided upon in stage 0.
Let $G_C$ denote this game (which was first analyzed by Vickers, 1985).

It is known that in the absence of delegation, the Stackelberg
market produces a higher total quantity than the Cournot market (Anderson and Engers, 1992). When delegation is introduced then: (i) In $G_S$ not all firms delegate; (ii) in $G_C$ all firms delegate. Hence a direct ranking of the Stackelberg and Cournot total market quantities under certain delegation is not obvious. Corollary 2 bellow provides this comparison. It also compares incentive rates and profiles across the two frameworks (in what follows, $Q^*_C$, $a^*_C$ and $u^*_C$ denote the equilibrium total market quantity, incentive rate and profit\footnote{Due to the symmetry of the model, all firms in the Cournot market choose the same incentive rate and have the same profit in equilibrium.} respectively under Cournot competition).

\vsp\noi \textbf{Corollary 2.} \emph{Consider the games $G_S$ and
$G_C$. The following hold.}

\vspace{0.2cm}\noi \emph{(i) For any $n \geq 2$, $Q^*_S>Q^*_C$.}

\vspace{0.2cm}\noi \emph{(ii) For any $n \geq 2$, $a^*_n>a^*_C$ and $a^*_C>a^*_i$, $i=1,2,\cdots,n-1$.}

\vspace{0.2cm}\noi \emph{(iii) For $n=2$, $u_2^*>u_C^*>u_1^*$;
for $n\geq 3$, then $u_C^*>u_i^*,$ for all $i$.}

\vspace{0.3cm}\noi \textbf{Proof.} Appears in the Appendix.

\vspace{0.2cm}

\noindent Corollary 2 shows an interesting relation: All firms in $G_S$, {\em except} for the last mover, choose lower incentive rates than firms in $G_C$. Nonetheless, regarding consumers, the Stackelberg market remains more efficient than the Cournot market as it results to a higher market quantity.
\vsp\noi In the absence of delegation, Anderson and Engers (1992) compared the profitability in the Stackelberg and Cournot models: for $n= 2$; the first (second) mover earns a higher (lower) profit than the Cournot duopolists; for $n \geq 3$, all Stackelberg firms earn a lower profit than the Cournot firms. When delegation is introduced and $n=2$, the second (first) mover earns a higher (lower) profit than the Cournot duopolists (this case is analyzed by Kopel and Loffler, 2008); for $n \geq 3$, all Stackelberg firms earn a lower profit than the Cournot firms (Corollary 2(iii)).

Let us, at this point, recall our assumption that each stage's choices are perfectly observable. The issue of imperfect observability in strategic games has been analyzed in a series of works. Katz (1991) demonstrated that if delegation choices are not observed by rivals then delegation has no value. Bagwell (1994) showed that observing the rival's action with noise destroys the impact of first mover's commitment. Vardy (2004) analyzed a sequential game where observing the first mover's choice is costly. He showed that being the first mover has no value, no matter how small the observation cost is. Other authors delivered more positive results: Fershtman and Kalai (1997) provided a framework where the value of delegation can be restored, provided there is a positive probability that the delegation contracts are accurately observed. van Damme and Hurkens (1997), Guth {\em et al.} (1998) and Maggi (1999) showed that commitment under imperfect observability has an impact on the outcome of the game if one allows for either mixed strategy equilibria (first two papers) or for private information on behalf of the first mover (last paper).

Contributing to the above discussion is not a goal (or an ambition) of the current paper. Just to provide some real-world facts we quote from Scalera and Zazzaro (2008):

"$\cdots$ the assumption of contract observability seems in some cases to be quite realistic. When firms compete to hire managers, it is likely that contractual clauses are publicly declared."

Further, the same present the argument that

"$\cdots$ in many countries, at least as regards quoted companies, firms are obliged by regulators to announce manager compensations to the market and this eases their commitment to the contracts signed with managers."\footnote{Katz (1991) questions the impact of this type of announcements as they refer to the actual payments of the managers and not to the rules that generate these payments.}

\section{Conclusions}
We analyzed strategic delegation in a Stackelberg model with an
arbitrary number of firms. We showed that the later a firm's manager
commits to a quantity, the higher his firm's profit. Delegation
improves the payoff of the late-movers and hurts early-movers. 
Namely, firms whose managers commit late (early) to a quantity 
end-up with a higher (lower) payoff compared to the non-delegation
regime. This is different from the case of delegation under Cournot
competition, where all firms are hurt by delegation.

Our paper has analyzed a framework with linear demand and cost functions.
Introducing a more general framework will allow us to examine 
the robustness of our results. Further the analysis of a market where incentive contracts are
imperfectly observed is of special interest.

\section*{Appendix}
\noi \textbf{Proof of Lemma 1.} (i) Notice that
$$\frac{\partial{u}_i(a_i,a_{-i})}{\partial{a}_i}>0\Leftrightarrow\frac{1}{2^i}(\frac{a}{2^n}+\sum_{j=1}^n\frac{c-a_j}{2^j}-c+a_i)
+(\frac{a}{2^n}+\sum_{j=1}^n\frac{c-a_j}{2^j}-c)(1-\frac{1}{2^i})>0$$

\vsp\noi or iff $$a_i<(\frac{a-c}{2^n}-\sum_{j\neq i}\frac{a_j}{2^j})\frac{2^i(2^i-2)}{2^{i+1}-2}$$

Clearly, for $i=1$, the derivative is negative and hence the equilibrium incentive rate that firm 1 chooses is $a_1^*=0$. Let now $i\geq 2.$ Then the reaction function of firm $i$ is given by

\begin{displaymath} a_i= \left\{
\begin{array}{ll}
 0,&\mbox{ if }\sum_{j\neq i}a_j/2^j\geq (a-c)/2^n,\\
\frac{2^i}{\sigma(i)}[(a-c)/2^n-\sum_{j\neq i}a_j/2^j],&\mbox{ if } \sum_{j\neq i}a_j/2^j<(a-c)/2^n .\\
\end{array}\right.\end{displaymath}

\vsp\noi where $\sigma(i)=(2^{i+1}-2)/(2^i-2)$, $i\geq 2$. We first notice that in any equilibrium of the
delegation game only firm 1's owners choose a zero incentive rate; all the remaining firms choose positive incentive rates. To see this, consider an outcome $a_i=0$ and $a_j>0, j=2,\cdots,
i-1,i+1,\cdots,n$, $j\neq i$. The market price is ${\ds{P=\frac{a}{2^n}+\sum_{i=1}^n\frac{c-a_i}{2^i}}}.$
Since $a_i=0$, we have $\sum_{j\neq i}a_j/2^j\geq (a-c)/2^n$. But then it is easy to show the last inequality implies
that the corresponding price would fall below the marginal cost $c$. Hence we restrict attention to the case
where $\sum_{j\neq i}a_j/2^j<(a-c)/2^n$ for all $i$. Then we have the system

\be\frac{1}{2^2} a_2+\frac{1}{2^3} a_3+\cdots+
\sigma(i)\frac{1}{2^i} a_i+\cdots+\frac{1}{2^n} a_n=\frac{a-c}{2^n},\hspace{0.2cm}i=2,3,...,n\ee
$\sigma(i)=\frac{2^{i+1}-2}{2^i-2}$. Using (5) the equations for firms $i$ and 2 we get,
\be a_i=\frac{\sigma(2)-1}{\sigma(i)-1} 2^{i-2} a_2\ee
and hence \be a_2=\frac{a-c}{2^{n-2}}(\sigma(2)+\sum^n_{i=3}\frac{\sigma(2)-1}{\sigma(i)-1})^{-1}\ee

\vsp\noi It is straightforward to show that ${\ds{\sigma(2)+\sum^n_{i=3}\frac{\sigma(2)-1}{\sigma(i)-1}=-2+2n
+2^{2-n}\equiv h(n)}}$. Using then (6) and (7), the solution for $a_i, i\geq 2,$ is ${\ds{a_i^*=D_i\frac{a-c}{2^n}\frac{1}{h(n)}}},$
where ${\ds{D_i=\frac{2^{i+1}}{\sigma(i)-1}.}}$

\vsp\noi(ii) Notice that $a_{i+1}^*>a_i^*$ if and only if
$D_{i+1}>D_i$ or $2^{i+2}/(\sigma(i+1)-1)> 2^{i+1}/(\sigma(i)-1)$,
which holds because $\sigma(i+1)<\sigma(i).$\qed

\vspace{0.2cm}

\noi \textbf{Proof of Corollary 1.} The equilibrium profit of the
$i$-th mover in $G_S$ is $u_i^*=(a-c)^2(1-2^{-i})/[2^{n-2}[h(n)]^2]$
whereas the profit of the same firm in a market without delegation
activities is $\bar{u}_i=2^{-i}(a-c)^2/2^n.$ Then, $u_i^*>\bar{u}_i$
if and only if $2^{2+i}>4+[h(n)]^2$. Let $r(i)=2^{2+i}$. It is easy
to show that $r(1)<4+[h(n)]^2<r(n)$; further, $r(i)$ is increasing
in $i$. Hence there exists a unique $i'(n)<n$ such that $u_i^*>
\bar{u}_i$ if and only if $i>i'(n)$.\qed

\vspace{0.2cm}

\noi \textbf{Proof of Corollary 2.}(i) Consider the last stage of
$G_C$. The quantity that the manager of firm $i$ chooses
is $$q_{iC}(a)=max\{(a-n(c-a_i)+\sum_{i\neq
j}(c-a_j))/(n+1),0\},\hspace{0.2cm} i=1,2,...,n$$ Equilibrium
delegation schemes are
$${\ds{a_i^{*}=a_{C}^*=\frac{n-1}{n^2+1}(a-c),\hspace{0.2cm}
i=1,2,\cdots,n}}$$ Hence, individual and total market quantities are
given respectively by $$q_{iC}^*=q_C^*=\frac{n(a-c)}{n^2+1},
\hspace{0.2cm}
Q_{C}^*=\frac{n^2(a-c)}{n^2+1},\hspace{0.2cm}i=1,2,\cdots,n$$ Recall
that market quantity in $G_S$ is
$${\ds{Q_{S}^*=(a-c)[1-2^{-n}+\frac{2n-4+2^{2-n}}{2^nh(n)}]}}$$ It is
then easy to show that $Q_S^*>Q_C^*$ if and only if
$(n-1)2^{1+n}+2-2n^2>0$ which holds.

\vspace{0.3cm}\noi (ii) Notice that $a_i^*>a_C^*$ iff
$2^{i+1}>4+[(n-1)2^nh(n)]/(n^2+1).$ Define the function
$w(i)=2^{i+1}$ and notice that
$w(n-1)<4+[(n-1)2^nh(n)]/(n^2+1)<w(n)$. Since $w(i)$ is strictly
increasing in $i$, we conclude that $a_i^*<a_C^*$ for
$i=1,2,\cdots,n-1$ and $a_n^*>a_C^*.$

\vspace{0.3cm}\noi (iii) The equilibrium profit of the $i$-th mover
in $G_S$ is ${\ds{u_i^*=\frac{4(1-2^{-i})}{2^n[h(n)]^2}(a-c)^2}}$
whereas the profit of each firm in $G_C$ is{\footnote{In the absence
of delegation, each Cournot firm earns $(a-c)^2/(n+1)^2$. Thus the
non-delegation regime is preferable by all Cournot firms over the
delegation regime.}} ${\ds{u_C^*=\frac{n}{(n^2+1)^2}(a-c)^2}}.$
Notice that $u_i^*>u_C^*$ if and only if
${\ds{4-2^{2-i}>\frac{n2^n(h(n))^2}{(n^2+1)^2}.}}$ Let $y(n)$ denote
the right part of the last inequality. For $n\geq 3$,
$y(n)>4>4-2^{2-i}$. On the other hand, if $n=2$,
$u_1^*=(a-c)^2/18<u_C^*=2(a-c)^2/25<u_2^*=(a-c)^2/12.$ \qed

\section*{References}

\noi Anderson S., Engers M., (1992), Stackelberg versus Cournot
oligopoly equilibrium, International Journal of Industrial
Organization, 10, 127-135.

\vspace{0.2cm}\noi Bagwell, K. (1995) Commitment and observability in games, Games Econ. Behav. 8, 271-280.

\vspace{0.2cm}\noi Boyer, M. and Moreaux, M.(1986), Perfect
competition as the limit of a hierarchical market game, Economics
Letters, 22, 115-118.

\vspace{0.2cm}\noi Fershtman C. and Judd, K.L. (1987), Equilibrium
incentives in oligopoly, American Economic Review, 77, 927-940.

\vspace{0.2cm}\noi Fershtman C. and Kalai, E. (1997), Unobserved delegation, Int. Econ. Rev. 38, 763-774.

\vspace{0.2cm}\noi Gal-Or E. (1985), First mover and second mover
advantages, International Economic Review, 26, 649-653.

\vspace{0.2cm}\noi Gal-Or E. (1987), First mover disadvantages
with private information, Review of Economic Studies, 54, 279-292.

\vspace{0.2cm}\noi Guth, W., Kirchsteige, G. and Ritzberger, K. (1998), Imperfectly observable commitments in $n$-player games, Games Econ. Behav. 23, 54-74.

\vspace{0.2cm}\noi Katz, M. L. (1991) Game-playing agents: Unobservable contracts as pre-commitments, Rand J. Econ. 22, 307-328.

\vspace{0.2cm}\noi Kopel M., and Loffler, C. (2008), Commitment,
first-mover and second-mover advantage, Journal of Economics 94,
143-166.

\vspace{0.2cm}\noi Liu Z. (2005), Stackelberg leadership with
demand uncertainty, Managerial and Decision Economics, 26, 345-350.

\vspace{0.2cm}\noi Maggi, G. (1999) The value of commitment with imperfect observability and private information, Rand J. Econ. 30(4), 555-574.

\vspace{0.2cm}\noi Pal D. and Sarkar, J. (2001), A Stackelberg
oligopoly with non-identical firms, Bulletin of Economic Research,
53, 127-134.

\vspace{0.2cm}\noi Scalera, D. and Zazzaro, A. (2008) Observable managerial incentives and spatial competition, Metroeconomica 59, 27-41.

\vspace{0.2cm}\noi Sklivas S. (1987), The strategic choice of
managerial incentives, Rand Journal of Economics, 18, 452-458.

\vspace{0.2cm}\noi van Damme, E. and Hurkens, S. (1997) Games with imperfectly observable commitment, Games Econ. Behav. 21, 282-308.

\vspace{0.2cm}\noi Vardy F. (2004), The value of commitment in
Stackelberg games with observation costs, Games and Economic
Behavior 49, 374-400.

\vspace{0.2cm}\noi Vickers J. (1985), Delegation and the theory
of the firm, Economic Journal 95, 138-147.

\end{document}